\documentclass[pra,fleqn,showpacs]{revtex4}

\usepackage{amsmath,epsfig}
\usepackage{dcolumn}
\begin{document}

\title{The Bethe Logarithm for resonant states. Antiprotonic helium.}
\author{Vladimir I.~Korobov}
\affiliation{Joint Institute for Nuclear Research\\
141980, Dubna, Russia}

\begin{abstract}
We develop a numerical method to calculate the Bethe logarithm for resonant states. We use the Complex Coordinate Rotation (CCR) formalism to describe resonances as time-independent Schr\"odinger solutions. To get a proper expression for the Bethe logarithm we apply the generalization of the second order perturbation theory to an isolated CCR eigenstate. Using the developed method we perform a systematic calculation of the Bethe logarithm for metastable states in the antiprotonic helium He$^+\bar{p}\,$ atoms with precision of 7--8 significant digits. We also recalculate the nonrelativistic energies with improved precision using CODATA10 recommended values of masses. Along with a complete set of corrections of $m\alpha^7$ order and the leading contributions of $m\alpha^8$ order that has allowed us to get new theoretical values for ro-vibrational transition frequencies for the He$^+\bar{p}$ atoms with uncertainty of 0.1--0.3 MHz.
\end{abstract}

\maketitle

\section{Introduction}

Precision spectroscopy of the antiprotonic helium is considered as one of the possible ways to improve the CODATA value for the atomic mass of an electron \cite{CODATA10,Hori11} assuming the validity of the CPT symmetry at this level of accuracy. Since the discovery of a long-lived fraction of antiprotons in helium \cite{first} and the first laser experiments \cite{Torii} a great progress in precision from ppm to ppb level has been achieved \cite{Hori,Hori06}. More details on this exotic system may be found in \cite{PhysRep,HayanoRep}.

On the other hand, it was shown that individual states of the antiprotonic helium may be treated numerically with high precision \cite{Kor_PRA96}. Despite the fact that these states appear in the continuum of the nonrelativistic Hamiltonian operator as resonances having antiprotons in nearly circular orbitals with total orbital angular momentum of an atom $L\sim30-36$, they allow to be calculated very precisely taking into account their resonant nature \cite{Kor03} together with many higher order (in powers of the fine structure constant $\alpha$) relativistic and radiative corrections \cite{Kor08}.

The major goal of the present work is to get a fractional precision of one part in $10^{10}$ for the theoretical transition frequencies, which should be compared with the CODATA10 \cite{CODATA10} uncertainty limits for the atomic mass $4.1\cdot10^{-10}$. To achieve that we need to solve two problems. The first one is to calculate the nonrelativistic Bethe logarithm for individual metastable states with accuracy of 7 significant digits going beyond the usual bound state formalism \cite{Kor03}. The second is to obtain the complete set of contributions of various corrections of $m\alpha^7$ order. The latter was carried out recently in \cite{PRA13,a7}. The former will be considered here below.

\section{Resonances and the Complex Coordinate Rotation approach}

To have a rigorous background for our calculations we need to give a brief outline of the Complex Coordinate Rotation (CCR) method \cite{Ho83} along with some basics for the perturbation theory for isolated resonant states. The Coulomb Hamiltonian for a system of point-like particles is analytic under dilatation transformations
\begin{equation}\label{dilatation}
\left(U(\theta)f\right)(\mathbf{r}) = e^{m\theta/2}f(e^\theta\mathbf{r}),
\qquad H(\theta)=U(\theta)HU^{-1}(\theta),
\end{equation}
for real $\theta$ and can be analytically continued to the complex plane. The Complex Coordinate Rotation method \cite{Ho83} "rotates" the coordinates of the dynamical system ($\theta=i\varphi$), $r_{ij}\rightarrow r_{ij} e^{i\varphi}$, where $\varphi$ is the parameter of the complex rotation. Under this transformation the Hamiltonian
changes as a function of $\varphi$
\begin{equation}
H_{\varphi} = T e^{-2 i \varphi} + V e^{-i \varphi},
\end{equation}
where $T$ and $V$ are the kinetic energy and Coulomb potential operators. The continuum spectrum of $H_{\varphi}$ is rotated on the complex plane around branch points ("thresholds") to "uncover" resonant poles situated on the unphysical sheet of the Reimann surface in accordance with the Augilar-Balslev-Combes theorem \cite{ABC}. The resonance energy is then determined by solving the complex eigenvalue problem for the "rotated" Hamiltonian
\begin{equation}
(H_{\varphi} - E)\Psi_{\varphi} = 0. \label{roteqn}
\end{equation}
The eigenfunction $\Psi_{\varphi}$ obtained from Eq.~(\ref{roteqn}), is square-integrable and the corresponding complex eigenvalue $E = E_r - i\Gamma/2$ defines the energy $E_r$ and the width of the resonance, $\Gamma$, the latter being related to the Auger rate as $\lambda_A = \Gamma/\hbar$.

The use of a finite set of $N$ basis functions reduces the problem (\ref{roteqn}) to the generalized algebraic complex eigenvalue problem
\begin{equation}
(A-\lambda B) x = 0, \label{gaevalp}
\end{equation}
where $A=\langle\Psi_{\varphi}|H_{\varphi}|\Psi_{\varphi}\rangle$ is the finite $N\times N$ matrix of the Hamiltonian in this basis, and $B$ is the matrix of overlap $B=\langle\Psi_{\varphi}|\Psi_{\varphi}\rangle$.

To evaluate the nonrelativistic Bethe logarithm for the CCR states a second-order perturbation theory is necessary. The relevant background is provided by the theorem \cite{Simon}.

\textbf{Theorem.\,} Let $H$ be a three-body Hamiltonian with Coulomb pairwise interaction, and $W(\theta)$ be a dilatation analytic perturbation. Let $E_0$ be an isolated simple resonance energy (discrete eigenvalue of $H(\theta)$). Then for $\beta$ small, there is exactly one eigenstate of $H(\theta)+\beta W(\theta)$ near $E_0$ and
\[
E(\beta)=E_0+a_1\beta+a_2\beta^2+\dots
\]
is analytic near $\beta=0$. In particular,
\begin{equation}
\begin{array}{@{}l}\displaystyle
a_1 = E'(0) =
   \left\langle\Psi^*_0(\theta)\left| W(\theta) \right|\Psi_0(\theta)\right\rangle,
\\[3mm]\displaystyle
a_2 = \sum_{n\ne0}
   \frac{
      \left\langle\Psi^*_0(\theta)\left| W(\theta) \right|\Psi_n(\theta)\right\rangle
      \left\langle\Psi^*_n(\theta)\left| W(\theta) \right|\Psi_0(\theta)\right\rangle}
         {E_0-E_n(\theta)}
\end{array}
\end{equation}
It is assumed that the wave functions are normalized as $\left\langle\Psi_\theta^*,\Psi_\theta\right\rangle=1$. Coefficients $a_1$, $a_2$, etc do not depend on $\theta$ if only branches uncover $E_0$ and its vicinity on the complex plane.

\section{Leading order radiative corrections and the Bethe logarithm}

The complete spin-independent contribution of orders $m\alpha^5$ and $m\alpha^5(m/M)$ for a one electron molecular-type system may be expressed by three terms: the one-loop self-energy correction, the transverse photon exchange term, and the vacuum polarization \cite{Pachucki,Yelkhovsky}.

The one-loop self-energy correction ($R_\infty\alpha^3$) has the following form:
\begin{equation}\label{SE}
\begin{array}{@{}l}\displaystyle
E_{se}^{(3)} = \alpha^3\frac{4}{3}
   \left[\ln{\frac{1}{\alpha^2}}-\beta(L,v)+\frac{5}{6}-\frac{3}{8}\right]
   \left\langle
      Z_{\rm He}\delta(\mathbf{r}_{\rm He}^{})+
      Z_{\bar{p}}\delta(\mathbf{r}_{\bar{p}})
   \right\rangle,
\end{array}
\end{equation}
where
\begin{equation}\label{BL}
\beta(L,v) =
   \frac{
   \left\langle
      \mathbf{J}(H\!-\!E_0)\ln\left((H\!-\!E_0)/R_\infty\right)\mathbf{J}
   \right\rangle}
   {\left\langle
      [\mathbf{J},[H,\,\mathbf{J}]]/2
   \right\rangle}
\end{equation}
is the nonrelativistic Bethe logarithm \cite{BS} for a bound state of the three-body system. Here $\mathbf{J}=\sum_a Z_a\mathbf{p}_a/m_a$ is the electric current density operator of the whole system. It is known that the Bethe logarithm is one of the most difficult quantities to evaluate numerically in atomic physics. So far, for the case of the antiprotonic helium it was calculated based on the closed-channel variational approximation for the initial wave function \cite{Kor03}. In this case a state may be considered as a "true" bound state. This approximation was limited in accuracy by four to six significant digits, and become unsatisfactory for present level theoretical estimates.

\begin{table}[t]
\begin{center}
\begin{tabular}{c@{\quad}c@{\quad}l@{\quad}l@{\quad}l@{\quad}l@{\quad}l@{\quad}l}
\hline\hline
state & $\Delta l$ & $~~~~~~~~~~E_{nr}$ & $~~~~\Gamma/2$ & $~~~~~{\bf p}^4_e$
 & $~~\delta({\bf r}_{_{\rm He}})$ & $~~\delta({\bf r}_{\bar p})$ & $~~\beta(n,l)$ \\
\hline\hline
(31,30) & 3 & $-$3.67977478748142(4) & $4.76010\cdot10^{-9}$& 26.070960 & 0.92622196& 0.12144043 & 4.578969(1) \\
(32,31) & 4 & $-$3.507635038808513(2)& $5.36\cdot10^{-13}$  & 28.308650 & 0.99382380& 0.11308041 & 4.560196(1) \\
(33,32) & 4 & $-$3.353757870683624(4)& $1.060\cdot10^{-12}$ & 30.718284 & 1.0664983 & 0.10445828 & 4.5416885(4)\\
(34,32) & 3 & $-$3.2276763794925(1)  & $2.7236\cdot10^{-9}$ & 34.530626 & 1.1808674 & 0.09255952 & 4.512829(1) \\
(34,33) & 4 & $-$3.216244238932181(1)& $1.38\cdot10^{-13}$  & 33.304865 & 1.1443963 & 0.09561357 & 4.523626(1) \\
(35,32) & 3 & $-$3.1166797957470(5)  & $6.97306\cdot10^{-8}$& 38.370061 & 1.2958621 & 0.08121154 & 4.488492(4) \\
(35,33) & 4 & $-$3.10538267542400(5) & $2.67\cdot10^{-12}$  & 37.278814 & 1.2635240 & 0.08387045 & 4.496653(1) \\
(35,34) & 5 & $-$3.0934669077893306  &         ---          & 36.069959 & 1.2275614 & 0.08659337 & 4.5061577(4)\\
(36,33) & 3 & $-$3.0079790935681(3)  & $2.9186\cdot10^{-9}$ & 41.233444 & 1.3819867 & 0.07291740 & 4.474194(2) \\
(36,34) & 4 & $-$2.996335447851055(3)& $2.66\cdot10^{-13}$  & 40.168790 & 1.3503395 & 0.07513623 & 4.4812666(4)\\
(37,34) & 4 & $-$2.91118093936496(5) & $2.60\cdot10^{-12}$  & 44.174191 & 1.4702684 & 0.06466985 & 4.4608414(4)\\
(37,35) & 5 & $-$2.8992821832621387(5)& $1.2\cdot10^{-15}$  & 43.186472 & 1.4409042 & 0.06644874 & 4.4667491(3)\\
(38,34) & 3 & $-$2.8365246011112(6)  & $1.6029\cdot10^{-9}$ & 48.000302 & 1.5848214 & 0.05532901 & 4.4441865(5)\\
(38,35) & 4 & $-$2.825146809449515(3)& $1.64\cdot10^{-13}$  & 47.185112 & 1.5605892 & 0.05662323 & 4.4484354(3)\\
(39,34) & 3 & $-$2.771011573490(1)   & $0.9920\cdot10^{-8}$ & 51.574881 & 1.6918639 & 0.04717053 & 4.430698(3) \\
(39,35) & 4 & $-$2.76023334548707(3) & $0.93\cdot10^{-12}$  & 50.925521 & 1.6725710 & 0.04806117 & 4.433733(1) \\
(40,35) & 4 & $-$2.70328321643503(5) & $1.91\cdot10^{-12}$  & 54.349323 & 1.7751252 & 0.04075701 & 4.421998(5) \\
(40,36) & 4 & $-$2.69262484981043(3) & $2.02\cdot10^{-12}$  & 53.823828 & 1.7594940 & 0.04122551 & 4.424271(1) \\
(41,35) & 3 & $-$2.6531667754306(4)  & $1.4400\cdot10^{-9}$ & 57.423461 & 1.8672689 & 0.03463431 & 4.412720(3) \\
\hline\hline
\end{tabular}
\end{center}
\caption{Multipolarities of the Auger transition $\Delta l$, nonrelativistic energies $E_{nr}$ (in a.u.), Auger widths $\Gamma$ (in a.u.), expectation values of operators: ${\bf p}^4_e$, $\delta({\bf r}_{_{\rm He}})$, and $\delta({\bf r}_{\bar p})$, and the Bethe logarithm values, $\beta(n,l)$, for the Auger states of $^4\mbox{He}^+{\bar p}$ atom.}
\label{He4}
\end{table}

Next term is the recoil correction of order $R_\infty\alpha^3(m/M)$ \cite{Pachucki,Yelkhovsky}:
\begin{equation}\label{recoil}
E_{recoil}^{(3)}=\sum_{i=1,2}\frac{Z_i\alpha^3}{M_i}
   \Biggl\{
      \frac{2}{3}\left(
         -\ln\alpha-4\beta(L,v)+\frac{31}{3}
      \right)\left\langle\delta(\mathbf{r}_i)\right\rangle
      -\frac{14}{3} Q(r_i)
   \Biggr\},
\end{equation}
where $\beta(L,v)$ is the same Bethe logarithm quantity as in Eq.~(\ref{SE}), $Q(r)$ is the so-called Araki-Sucher term \cite{AS}:
\[
Q(r) = \lim_{\rho \to 0} \left\langle
            \frac{\Theta(r - \rho)}{ 4\pi r^3 }
      + (\ln \rho + \gamma_E)\delta(\mathbf{r}) \right\rangle.
\]
And the last one is the one-loop vacuum polarization:
\begin{equation}\label{VP}
E_{vp}^{(3)} = \frac{4\alpha^3}{3}
   \left[-\frac{1}{5}\right]
   \Bigl\langle
      Z_{\rm He}\delta(\mathbf{r}_{\rm He}^{})+
      Z_{\bar{p}}\delta(\mathbf{r}_{\bar{p}})
   \Bigr\rangle.
\end{equation}
The two quantities: the $Q(r)$ term and the mean value the $\delta$-function operator, which appear in Eqs.~(\ref{recoil}) and (\ref{VP}), can be easily evaluated for a CCR wave function of a stationary solution for a metastable state. In what follows in this section we explain, how calculation of the nonrelativistic Bethe logarithm for a bound state may be extended to resonant states using the Complex Coordinate Rotation formalism.

The Bethe logarithm for a "rotated" state in the coordinate system rotated by the same angle $\varphi$ ($r_{ij}\rightarrow r_{ij} e^{i\varphi}$) is expressed:
\begin{equation}\label{BL_rot}
\beta(L,v) =
   \frac{
   \left\langle
      \mathbf{J}_\varphi
      (H_\varphi\!-\!E_0)\ln\left((H_\varphi\!-\!E_0)/R_\infty\right)
      \mathbf{J}_\varphi
   \right\rangle}
   {\left\langle
      [\mathbf{J}_\varphi,[H_\varphi,\,\mathbf{J}_\varphi]]/2
   \right\rangle}
\end{equation}
where $H_{\varphi}$ and $\mathbf{J}_{\varphi}$ are the rotated operators of the Hamiltonian and the charge current density. It is better to rewrite this quantity in an equivalent form as an integration over the virtual photon energy $k$:
\begin{equation}
\beta(L,v) =
   \frac{\displaystyle
   \int_0^{E_h}\,k\,dk
   \left\langle
      \mathbf{J}_\varphi
      \left(\frac{1}{E_0\!-\!H_\varphi\!-\!k}\!+\!\frac{1}{k}\right)
      \mathbf{J}_\varphi
   \right\rangle
   +\int_{E_h}^\infty\,\frac{dk}{k}\,
   \left\langle
      \mathbf{J}_\varphi\frac{(E_0\!-\!H_\varphi)^2}{E_0\!-\!H_\varphi\!-\!k}\mathbf{J}_\varphi
   \right\rangle}{
   \bigl\langle
      \left[\mathbf{J}_\varphi\left[H_\varphi,\mathbf{J}_\varphi\right]\right]
   \bigr\rangle/2}\>.
\end{equation}
Its integrand may be expressed via a basic function $J(k)$, the contribution of the second order perturbation of a virtual photon emission and absorbtion,
\begin{equation}\label{Jk}
J(k) =
   \left\langle
      \mathbf{J}_\varphi\left(E_0\!-\!H_\varphi\!-\!k\right)^{-1}\mathbf{J}_\varphi
   \right\rangle.
\end{equation}
We neglect retardation as is usual for the nonrelativistic Bethe logarithm calculations \cite{BS}. It is worthy to note that $J(k)$ does not depend on $\varphi$ for a complete (infinite) basis set, and thus the final value for $\beta(L,v)$ will not depend on the "unphysical" parameter --- the rotational angle. Meanwhile, the number itself should be complex with the imaginary part being the radiative correction contribution to the Auger decay rate.

The actual calculation of the Bethe logarithm for the metastable states of the antiprotonic helium is performed as a straightforward generalization of the numerical scheme derived in \cite{Kor12BL}, and all the technical details may be found there.

\begin{table}[t]
\begin{center}
\begin{tabular}{c@{\quad}c@{\quad}l@{\quad}l@{\quad}l@{\quad}l@{\quad}l@{\quad}l}
\hline\hline
state & $\Delta l$ & $~~~~~~~E_{nr}$ & $~~\Gamma/2$ & $~~~{\bf p}^4_e$
& $\delta({\bf r}_{_{\rm He}})$ & $\delta({\bf r}_{\bar p})$ & $\beta(n,l)$ \\
\hline\hline
(31,30) & 3 & $-$3.5073727202819(5)  & $3.3424\cdot10^{-9}$ & 28.309519 & 0.99368837& 0.11287882 & 4.559722(1) \\
(32,31) & 4 & $-$3.348832173150003(2)& $5.169\cdot10^{-12}$ & 30.803393 & 1.0689407 & 0.10401090 & 4.540686(1) \\
(33,31) & 3 & $-$3.2195072516355(1)  & $8.2761\cdot10^{-9}$ & 34.744079 & 1.1871602 & 0.09174227 & 4.511062(1) \\
(33,32) & 4 & $-$3.20767231244689(1) & $7.8\cdot10^{-13}$   & 33.484950 & 1.1497243 & 0.09489883 & 4.522121(1) \\
(34,31) & 3 & $-$3.1061288628903(2)  & $7.925\cdot10^{-10}$ & 38.697601 & 1.3055346 & 0.08005938 & 4.486207(2) \\
(34,32) & 4 & $-$3.09445096699891(2) & $1.709\cdot10^{-11}$ & 37.595341 & 1.2729446 & 0.08281128 & 4.4945424(5)\\
(34,33) & 5 & $-$3.082114107332030(1)&       ---            & 36.355772 & 1.2360854 & 0.08559822 & 4.5041924(5)\\
(35,32) & 3 & $-$2.9954043586889(1)  & $8.1608\cdot10^{-9}$ & 41.676373 & 1.3951949 & 0.07158519 & 4.471761(1) \\
(35,33) & 4 & $-$2.983373123874257(5)& $1.303\cdot10^{-12}$ & 40.593960 & 1.3630401 & 0.07383439 & 4.4788451(4)\\
(36,32) & 3 & $-$2.9087979813554(5)  & $5.7466\cdot10^{-9}$ & 45.621175 & 1.5132762 & 0.06138275 & 4.452989(2) \\
(36,33) & 4 & $-$2.89719228821683(3) & $2.915\cdot10^{-10}$ & 44.720654 & 1.4866058 & 0.06315237 & 4.4582350(5)\\
(36,34) & 5 & $-$2.88491261972020(1) &       ---            & 43.723769 & 1.4569802 & 0.06491865 & 4.4640761(4)\\
(37,33) & 3 & $-$2.8219630311214(2)  & $4.2678\cdot10^{-9}$ & 48.642644 & 1.6040396 & 0.05369086 & 4.441533(3) \\
(37,34) & 4 & $-$2.81026108564305(5) & $7.6\cdot10^{-13}$   & 47.831121 & 1.5799277 & 0.05494800 & 4.4456893(6)\\
(38,33) & 3 & $-$2.756217741055(2)   & $3.4239\cdot10^{-8}$ & 52.279701 & 1.7129557 & 0.04549272 & 4.428246(1) \\
(38,34) & 4 & $-$2.74517414926844(2) & $3.90\cdot10^{-12}$  & 51.647733 & 1.6942064 & 0.04633993 & 4.4310677(4)\\
(39,34) & 4 & $-$2.688292963759(2)   & $1.130\cdot10^{-9}$  & 55.114275 & 1.7980611 & 0.03907099 & 4.419510(4) \\
(40,35) & 4 & $-$2.62832405152957(1) & $6.7\cdot10^{-13}$   & 57.840699 & 1.8798956 & 0.03318082 & 4.4118712(5)\\
\hline\hline
\end{tabular}
\end{center}
\caption{Multipolarities of the Auger transition $\Delta l$, nonrelativistic energies $E_{nr}$ (in a.u.), Auger widths $\Gamma$ (in a.u.), expectation values of operators: ${\bf p}^4_e$, $\delta({\bf r}_{_{\rm He}})$, and $\delta({\bf r}_{\bar p})$, and the Bethe logarithm values, $\beta(n,l)$, for the Auger states of $^3\mbox{He}^+{\bar p}$ atom.}
\label{He3}
\end{table}

\section{The variational wave function}

In our CCR calculations, the initial quasi-bound state of the antiprotonic helium atom as well as the intermediate states, which appears in the second order perturbation calculations of the Bethe logarithm, are taken in the form~\cite{var00},
\begin{equation}\label{eq:exp}
  \Psi_L(l_1,l_2)=
    \sum_{i=1}^{\infty}
        \left\{
  U_i\,\mbox{Re}\left[e^{-\alpha_iR-\beta_ir_1-\gamma_ir_2}\right]
  +W_i\,\mbox{Im}\left[e^{-\alpha_iR-\beta_ir_1-\gamma_ir_2}\right]
        \right\}
        \mathcal{Y}^{l_1,l_2}_{LM}(\hat{\mathbf{R}},\hat{\mathbf{r}}_1)\,,
\end{equation}
where $\mathcal{Y}^{l_1,l_2}_{LM}(\hat{\mathbf{R}},\hat{\mathbf{r}}_1)$ are the solid bipolar harmonics as defined in~\cite{Var88}, $L$ is the total orbital angular momentum of a state. The initial states have normal spatial parity: $\pi=(-1)^L$. Complex parameters $\alpha_i$, $\beta_i$ and $\gamma_i$ are generated in a quasi-random manner:
\begin{equation}
  \mbox{Re}[\alpha_i]=
  \left[\left\lfloor\frac{1}{2}i(i+1)\sqrt{p_{\alpha}}\right\rfloor(A_2-A_1)+A_1\right]\,,
\qquad
  \mbox{Im}[\alpha_i]=
  \left[\left\lfloor\frac{1}{2}i(i+1)\sqrt{q_{\alpha}}\right\rfloor(A'_2-A'_1)+A'_1\right]\,,
\end{equation}
$\lfloor{x}\rfloor$ designates the fractional part of $x$, $p_{\alpha}$ and $q_{\alpha}$ are some prime numbers, $[A_1,A_2]$ and $[A'_1,A'_2]$ are real variational intervals, which need to be optimized subjecting the "minimax" principle of the Rayleigh-Ritz variational method. Parameters $\beta_i$ and $\gamma_i$ are obtained in a similar way.

The intermediate states span over $L'=L,L\pm1$ with the spatial parity $\pi=-(-1)^L$, where $L$ is a total orbital angular momentum of the initial quasi-bound state. The basis set of intermediate states is composed of a regular part and two extra short-distance trial functions (for $\mathbf{r}_i\to0$, $i=1,2$) with exponentially growing parameters (see details in \cite{Kor12BL}). To keep the required numerical stability the quadruple and sextuple precision arithmetics have been used.

\section{Results}

The results of numerical calculations are presented in Tables \ref{He4} and \ref{He3}. The nonrelativistic energies and widths were recalculated with improved precision and using the CODATA10 recommended values for physical constants \cite{CODATA10}. The basis sets for these variational CCR calculations were taken up to $N=7000$ basis functions. In the Tables we also present new data for the expectation values of the ${\bf p}^4_e$, $\delta({\bf r}_{_{\rm He}})$, and $\delta({\bf r}_{\bar p})$ operators. These numbers are of particular importance for evaluating the leading order relativistic corrections ($m\alpha^4$) with precision better than 100 kHz in ro-vibrational transition frequencies. The last column contains data of the CCR calculations of the Bethe logarithm, which are our main result of this work. Only the real part of $\beta(L,v)$ is shown. We estimate that the values presented have precision of 7-8 significant digits. It allows to claim that the uncertainty arising in the leading order radiative contribution $m\alpha^5$ is now below 100 kHz.

In Table \ref{results} a list of transition frequencies of spectroscopic interest both for $^4\mbox{He}^+\bar{p}\,$ and $^3\mbox{He}^+\bar{p}\,$ atoms are collected. The theoretical data contains a complete set of contributions up to $m\alpha^7$ order and the leading contributions of the $m\alpha^8$ order \cite{a7}. The error bars indicate mainly the uncertainty, which is caused by the numerical inaccuracy in the one-loop self-energy calculations. The whole budget of the contributions for the $(36,34)\to(34,32)$ transition of the $^4\mbox{He}^+\bar{p}$ atom, and a total list of the corrections, which were included, are discussed in detail in \cite{a7}. The last column gives a comparison with the best available experimental measurements for these transitions.

In conclusion, the results of the calculations presented here allows us to infer the electron-to-(anti)proton mass ratio from comparison of theoretical data of Table \ref{results} and future improved experimental measurements with the ultimate relative precision of about $10^{-10}$. That is about an order of magnitude more precise than the CODATA recommended value for the atomic mass of an electron.

\begin{table*}[t]
\begin{center}
\begin{tabular}{r@{\hspace{3mm}}c@{\hspace{7mm}}d@{\hspace{8mm}}r@{}l}
\hline\hline
\vrule width 0pt height 10pt
&
\multicolumn{1}{c}{transition~~~~~~} & \multicolumn{1}{c}{theory~~~} &
\multicolumn{2}{c}{experiment}\\
\hline
\vrule width 0pt height 11pt
$^4\mathrm{He}^+\bar{p}$
&$(32,31)\to(31,30)$ & 1\,132\,609\,223.8(2) & 1\,132\,609\,209&(15) \\
&$(34,33)\to(35,32)$ &    655\,062\,102.2(2) &                       \\
&$(35,33)\to(34,32)$ &    804\,633\,058.3(1) &    804\,633\,059&(8)~ \\
&$(36,34)\to(35,33)$ &    717\,474\,002.0(2) &    717\,474\,004&(10) \\
&$(37,34)\to(36,33)$ &    636\,878\,152.1(1) &    636\,878\,139&(8)~ \\
&$(37,35)\to(38,34)$ &    412\,885\,132.7(2) &    412\,885\,132&(4)~ \\
&$(38,35)\to(39,34)$ &    356\,155\,990.8(4) &                       \\
&$(39,35)\to(38,34)$ &    501\,948\,755.1(2) &    501\,948\,752&(4)~ \\
&$(40,35)\to(39,34)$ &    445\,608\,572.4(4) &    445\,608\,558&(6)~ \\
\vrule width0pt height 10.5pt depth4pt
&$(33,32)\to(31,30)$ & 2\,145\,054\,858.1(2) & 2\,145\,054\,858&(5)~ \\
&$(36,34)\to(34,32)$ & 1\,522\,107\,060.3(2) & 1\,522\,107\,062&(4)~ \\
\hline
\vrule width 0pt height 11pt
$^3\mathrm{He}^+\bar{p}$
&$(32,31)\to(31,30)$ & 1\,043\,128\,580.4(2) & 1\,043\,128\,609&(13) \\
&$(34,32)\to(33,31)$ &    822\,809\,172.2(3) &    822\,809\,190&(12) \\
&$(35,33)\to(34,32)$ &    730\,833\,930.2(1) &                       \\
&$(36,33)\to(35,32)$ &    646\,180\,412.6(2) &    646\,180\,434&(12) \\
&$(36,34)\to(37,33)$ &    414\,147\,509.3(3) &    414\,147\,508&(4)  \\
&$(38,34)\to(37,33)$ &    505\,222\,281.1(3) &    505\,222\,296&(8)  \\
\vrule width0pt height 10.5pt depth4pt
&$(35,33)\to(33,31)$ & 1\,553\,643\,102.4(3) & 1\,553\,643\,100&(7) \\
\hline\hline
\end{tabular}
\end{center}
\caption{Theoretical predictions to transition frequencies $\nu$ (in MHz)
between metastable states, and comparison with the latest experiment
\cite{Hori06}. Calculations are performed with CODATA10 recommended
values.} \label{results}
\end{table*}

\section*{Acknowledgements}

The work has been supported by the Russian Foundation for Basic Research under a grant No.~12-02-00417-a and by the Heisenberg-Landau program of BLTP JINR, which is gratefully acknowledged.

\end{document}